\documentclass[11pt]{article}
\usepackage{graphicx}
\usepackage{epstopdf}
\usepackage{amssymb,bm}
\usepackage{bbm}
\usepackage{amsmath}
\usepackage{amsfonts}
\usepackage[margin=1.0in]{geometry}
\usepackage{cite}
\usepackage[shortlabels]{enumitem}
\usepackage{comment}

\usepackage{color}

\DeclareMathOperator{\E}{\mathbbmss{E}}

\DeclareMathOperator{\Hbb}{\mathbb{H}}
\allowdisplaybreaks

\usepackage{amsthm}
\newtheorem{theorem}{Theorem}
\newtheorem*{theorem*}{Theorem}
\newtheorem{lemma}{Lemma}

\theoremstyle{definition}  
\newtheorem{definition}{Definition}
\newtheorem{remark}{Remark}

\usepackage{chngcntr}
\counterwithout{theorem}{section}
\counterwithout{definition}{section}
\counterwithout{lemma}{section}
\counterwithout{remark}{section}
\counterwithout{assumption}{section}
\counterwithout{proposition}{section}
\counterwithout{corollary}{section}
\counterwithout{claim}{section}

\begin{document}
\title{Optimality of Treating Inter-Cell Interference as Noise Under Finite Precision CSIT
\footnotetext{The authors are with the Faculty of Electrical Engineering and Computer Science, Technische Universit\"{a}t Berlin, 10587 Berlin, Germany (e-mail: h.joudeh@tu-berlin.de; caire@tu-berlin.de).
This work was partially supported by the European Research Council (ERC) under the ERC Advanced Grant N. 789190, CARENET.}}
\author{Hamdi~Joudeh and Giuseppe~Caire}
\date{}
\maketitle
\begin{abstract}
In this work, we study the  generalized degrees-of-freedom (GDoF) of downlink and uplink cellular networks, modeled as Gaussian interfering broadcast channels (IBC) and  Gaussian interfering multiple access channels (IMAC), respectively.
We focus on regimes of low inter-cell interference, where single-cell transmission with power control and treating inter-cell interference as noise (mc-TIN) is GDoF optimal.
Recent works have identified two relevant regimes in this context: one in which the  GDoF region achieved through mc-TIN for both the IBC and IMAC is a convex polyhedron without the need for time-sharing (mc-CTIN regime), and a smaller (sub)regime where mc-TIN is GDoF optimal for both the IBC and IMAC (mc-TIN regime).
In this work, we extend the mc-TIN framework to cellular scenarios where channel state information at the transmitters (CSIT) is limited to finite precision.
We show that in this case, the GDoF optimality of mc-TIN extends to the entire mc-CTIN regime, where GDoF benefits due to interference alignment (IA) are lost. 
Our result constitutes yet another successful application of robust outer bounds based on the aligned images (AI) approach.
\end{abstract}
\newpage
\section{Introduction}
\label{sec:introduction}
In the past one-and-a-half decade, significant progress has been made on the problem of characterizing and 
understanding the information-theoretic capacity limits of wireless communication networks.
Most such progress has been made through taking few steps away from exact (and intractable) capacity limits, 
and instead pursuing tractable characterizations based on the degrees-of-freedom (DoF) and generalized degrees-of-freedom (GDoF) approximations \cite{Cadambe2008,Etkin2008,Jafar2010}.

Initial DoF and GDoF studies focused on idealized settings, where the availability of prefect channel state information at the transmitters (CSIT) is assumed. This has given rise to novel achievability schemes, most notably those based on interference alignment (IA) \cite{Jafar2011}.
In practical scenarios, however, imperfections in channel knowledge are inevitable, hence raising concerns about the adequacy of schemes that rely on highly accurate CSIT.
Such concerns have been recently confirmed by a new class of robust information-theoretic outer bounds, known as aligned images (AI) bounds, which show the complete loss of DoF benefits due to zero forcing (ZF) in multi-antenna broadcast channels (MISO-BC), as well as GDoF benefits due to IA in symmetric interference channels (IC), when  CSIT is limited to finite precision \cite{Davoodi2016,Davoodi2017a}.

The recent insights into the fragility of schemes that rely on highly precise CSIT, enabled by AI bounds, have given new vitality to robust schemes that only require coarse channel knowledge, 
including treating interference as noise, rate-splitting and layered superposition, as seen through a number of studies  \cite{Geng2015,Yi2016,Yi2020,Joudeh2020,Davoodi2019}. 
These works seek to identify operational regimes in which such robust (and simple)
schemes are DoF or GDoF optimal in various settings and networks.

Most relevant to this paper are studies that focus on the GDoF optimality of  power control and treating interference as Gaussian noise (in short, TIN).
A  breakthrough in this direction is due to Geng et al.  \cite{Geng2015}, who identified a wide regime of channel strengths in which TIN achieves the whole GDoF region of the $K$-user IC, known as the TIN regime.
In \cite{Yi2016}, Yi and Caire identified a broader regime, known as the convex-TIN (CTIN) regime,
in which the TIN achievable (TINA) GDoF region is a convex polyhedron without the need for time-sharing.
While \cite{Geng2015} and \cite{Yi2016} make no implicit assumptions regarding the availability of CSIT, the robustness of TIN renders these results valid under finite precision CSIT.
More recently, Chan et al. \cite{Chan2019} employed AI bounds to show that by explicitly 
limiting CSIT to finite precision, the GDoF optimality of TIN is extended to the CTIN regime, where 
GDoF benefits due to IA disappear. 
In this paper, we generalized this recent observation to cellular networks.

We are primarily interested in the multi-cell TIN (mc-TIN) framework introduced in \cite{Joudeh2019a} and \cite{Joudeh2019b} for uplink and downlink cellular networks, modeled as interfering multiple access channels (IMAC) and interfering broadcast channels (IBC), respectively.
In mc-TIN schemes, power control and single-cell transmission is used in each cell, while all inter-cell interference is treated as Gaussian noise.
A mc-TIN regime in which mc-TIN achieves the entire GDoF region, in both the IBC and IMAC, has been identified in 
\cite{Joudeh2019a,Joudeh2019b}; as well as a mc-CTIN regime, where the corresponding mc-TINA GDoF regions 
are convex without the need for time sharing.
In the main result of this paper (Theorem \ref{theorem:CTIN_optimality}, Section \ref{sec:main_result}), we show that the GDoF optimality of mc-TIN extends to the entire mc-CTIN regime in both the IBC and IMAC, under  finite  precision CSIT.
This settles the GDoF regions of the IBC and IMAC in the mc-CTIN regime under finite precision CSIT, and shows that 
benefits due to the multi-cell IA (mc-IA) schemes in \cite{Suh2011,Suh2008}
are entirely lost in this case.

\emph{Notation:}
For positive integers $z_{1}$ and $z_{2}$ where $z_{1} \leq z_{2}$, the sets $\{1,2,\ldots,z_{1}\}$ and $\{z_{1},z_{1}+1,\ldots,z_{2}\}$ are denoted by
$\langle z_{1} \rangle$ and $\langle z_{1}:z_{2}\rangle$, respectively.
For any real number  $a \in \mathbb{R}$, we have $(a)^{+} = \max\{0,a\}$.
Bold symbols  denote tuples, e.g. $\mathbf{a} = (a_{1},\ldots,a_{Z})$ 
and $\mathbf{A} = (A_{1},\ldots,A_{Z})$, while 
calligraphic symbols denote sets, e.g. $\mathcal{A} = \{ a_{1},\ldots,a_{Z} \}$.
For a sequence of superscripted variables $W^{[1]}, \ldots , W^{[K]}$, we use  
$W^{[1:K]}$ as a shorthand notation for the set  $\{ W^{[1]}, \ldots , W^{[K]} \}$.
\section{System Model}
Consider a $K$-cell cellular network in which each cell $k$, where $k \in \langle K \rangle$, comprises a base station denoted by BS-$k$ and $L_{k}$ user equipments, each denoted by UE-$(l_{k},k)$, where $l_{k} \in \langle L_{k} \rangle$.
The set of tuples corresponding to all UEs in the networks is given by
$\mathcal{K} \triangleq \left\{(l_{k},k) : l_{k} \in \langle L_{k} \rangle, k \in \langle K \rangle  \right\}$.
\subsection{IBC and Dual IMAC}
When operating in the downlink mode, we assume that the above cellular network is modeled by a Gaussian IBC.
The input-output relationship at the $t$-th channel use is given by
\begin{equation}
\label{eq:IBC_system model}
Y_{k}^{[l_{k}]}(t)
 = \sum_{i = 1}^{K} \bar{P}^{\alpha_{ki}^{[l_{k}]}}  G_{ki}^{[l_{k}]}(t) X_{i}(t)
+ Z_{k}^{[l_{k}]}(t).
\end{equation}
In the above, $ X_{i}(t) , Y_{k}^{[l_{k}]}(t), Z_{k}^{[l_{k}]}(t) \in \mathbb{C}$ are, respectively, 
the symbol transmitted by BS-$i$, the symbol received by UE-$(l_{k},k)$, and the zero-mean unit-variance additive white Gaussian noise (AWGN) at UE-$(l_{k},k)$.
The signal transmitted by BS-$i$ is subject to the unit average power constraint 
$\frac{1}{T}\sum_{t=1}^{T}\E \big[\big|X_{i}(t)\big|^{2}\big] \leq 1$,
where $T$ is the duration of the communication.
For GDoF purposes, we define $\bar{P} \triangleq  \sqrt{P}$, where $P$ is a nominal power parameter
that approaches infinity in the GDoF limit.
On the other hand, ${\alpha_{ki}^{[l_{k}]}} > 0$ is the channel strength parameter between BS-$i$ and UE-$(l_{k},k)$,
while $G_{ki}^{[l_{k}]}(t) \in \mathbb{C}$ is the corresponding channel fading coefficient.

When operating in the uplink mode, the roles of the transmitters and receivers in the IBC are reverse, from which we obtain a dual IMAC with an input-output model given by
\begin{equation}
\label{eq:IMAC_system model}
Y_{i}(t)  = \sum_{k = 1}^{K} \sum_{l_{k} = 1}^{L_{k}}  \bar{P}^{\alpha_{ki}^{[l_{k}]}}G_{ki}^{[l_{k}]} (t) X_{k}^{[l_{k}]}(t)
+ Z_{i}(t).
\end{equation}
During the  $t$-th channel use, $ X_{k}^{[l_{k}]}(t), Y_{i}(t), Z_{i}(t) \in \mathbb{C}$ are, respectively, the symbol transmitted by UE-$(l_{k},k)$, the symbol received by BS-$i$, and the zero-mean unit-variance AWGN at BS-$i$.
The signal transmitted by UE-$(l_{k},k)$ is subject to the unit average power constraint 
$\frac{1}{T}\sum_{t=1}^{T}\E \big[\big|X_{k}^{[l_{k}]}(t)\big|^{2}\big] \leq 1$.
\begin{remark}
\label{remark:strength_order}
Without loss of generality, we assume that users in each cell are in an ascending order with respect to their direct link strength levels (or SNRs).
That is:
\begin{equation}
\label{eq:strength_order}
\alpha_{kk}^{[1]}  \leq  \alpha_{kk}^{[2]} \leq \cdots \leq \alpha_{kk}^{[L_{k}]}, \ \forall k \in \langle K \rangle.
\end{equation}
\end{remark}
\subsection{Finite Precision CSIT}
The channel strengths ${\alpha_{ki}^{[l_{k}]}}$ are perfectly known to both the transmitters and receivers. 
The channel coefficients $G_{ki}^{[l_{k}]}(t)$, however, are perfectly known to the receivers 
but only available up to finite precision at the transmitters. It is worth recalling that by transmitters, 
we are referring to the BSs in the IBC, and the UEs in the IMAC.
The finite precision CSIT model implies that from the transmitters' points of view,  the joint and conditional probability density 
functions of the channel coefficients exist, and the peak values of these densities are bounded by a constant which is independent of $P$.
As a result, transmitted codewords may only depend on the distribution of the coefficients $G_{ki}^{[l_{k}]}(t)$
and are independent of the exact realizations \cite{Davoodi2016,Davoodi2017a}.
\subsection{Messages, rates and GDoF}
Messages, achievable rates,  capacity regions and GDoF regions, for both the IBC and IMAC, 
are all defined in  a standard manner, see, e.g.  \cite{Joudeh2019a,Joudeh2019b}.
A rate tuple is denoted by $\mathbf{R}= \big(R_{k}^{[l_{k}]}: (l_{k} , k) \in \mathcal{K}\big)$,
while a GDoF tuple is denote by $\mathbf{d}= \big(d_{k}^{[l_{k}]}: (l_{k}, k) \in \mathcal{K}\big)$.
The GDoF regions for the IBC and dual IMAC are, respectively, denoted by 
$\mathcal{D}^{\mathrm{IBC}}$  and $\mathcal{D}^{\mathrm{IMAC}}$.
\section{Multi-Cell TIN}
We consider mc-TIN as defined in  \cite{Joudeh2019a,Joudeh2019b}, where 
a  single-cell-type scheme is employed in each cell, while 
all inter-cell interference is treated as Gaussian noise.

\emph{IBC:} 
Each BS  employs Gaussian superposition coding with power control, 
while each UE employs successive decoding and treats all inter-cell interference as noise.
We assume a fixed ascending decoding order, where each UE-$(l_{k},k)$ successively decodes the signals intended to
the UEs  indexed by $(1,k), (2,k) ,\ldots, (l_{k},k)$, in that order.
We denote the achievable GDoF region by $\mathcal{D}_{\mathrm{TINA}}^{\mathrm{IBC}}$.

\emph{IMAC:} 
Each UE-$(l_{k},k)$ employs Gaussian coding with power control, while each BS-$k$ successively 
decodes its in-cell signals while treating inter-cell interference as noise, with a descending decoding order of $(L_{k},k), (L_{k} - 1,k) ,\ldots, (1,k)$.
We denote the achievable region by $\mathcal{D}_{\mathrm{TINA}}^{\mathrm{IMAC}}$.

As time-sharing is not allowed in the above-described mc-TIN schemes, 
$\mathcal{D}_{\mathrm{TINA}}^{\mathrm{IBC}}$ and $\mathcal{D}_{\mathrm{TINA}}^{\mathrm{IMAC}}$ are not convex in general. 
From the uplink-downlink duality in \cite{Joudeh2019b}, we know that 
\begin{equation}
\mathcal{D}_{\mathrm{TINA}}^{\mathrm{IBC}} = \mathcal{D}_{\mathrm{TINA}}^{\mathrm{IMAC}} = 
\mathcal{D}_{\mathrm{TINA}}^{\mathrm{mc}}
\end{equation}
where we use $\mathcal{D}_{\mathrm{TINA}}^{\mathrm{mc}}$  to denote the multi-cell TINA region, for both the IBC and its dual IMAC.
\begin{remark}
The above ascending and descending decoding orders used in the mc-TIN schemes for the IBC and IMAC, respectively, 
are referred to as the natural orders \cite{Joudeh2019b}.
From a GDoF region viewpoint, these natural decoding orders are optimal in the absence of inter-cell interference, i.e.  
$\alpha^{[l_{i}]}_{ij} = 0$ for all $l_{i},i,j$, where $i \neq j$.
The presence of inter-cell interference, however, may change the relative strengths of different users, rendering such orders suboptimal in general---see, e.g., \cite[Fig. 2(c)]{Joudeh2019a}.
Nevertheless, in the regimes of interest, these decoding orders suffice as 
seen further on. 
\end{remark}
\subsection{Useful Definitions}
Here we present some definitions, which are instrumental to the formulation of our results,
and highlight their significance.
\begin{definition}
\textbf{(Cycle).}
A cycle of length $M$ is an ordered sequence of $M$ cells, given by
\begin{equation}
\sigma = (k_{1} \rightarrow k_{2}  \rightarrow \ldots \rightarrow k_{M} )
\end{equation}
where the participating set of cells  $\{k_{1},k_{2},\ldots,k_{M}\} \subseteq \langle K \rangle$ is denoted by $\{ \sigma \}$.
We use $\sigma (m)$ to denote  cell $k_{m}$, and $|\sigma|$ to denote $M$.
A modulo-$M$ operation is implicitly used on indices, e.g. $\sigma (M + 1) = \sigma(1)$ and $\sigma (0) = \sigma(M)$.
We assume that $|\sigma| \geq 2$.  The set of all cycles is denoted by  $\Sigma$.
\end{definition}
The significance of cycles is that they are  useful for describing polyhedral-TIN regions---essential building blocks for characterizing TINA GDoF regions, see, e.g., \cite{Geng2015,Chan2019,Joudeh2019a,Joudeh2019b}.
We now define a polyhedral-mc-TIN region of interest. 
\begin{definition}
\label{def:polyhedral_region}
\textbf{(polyhedral-mc-TIN region).}
This is described by all 
GDoF tuples $\mathbf{d} \in \mathbb{R}_{+}^{|\mathcal{K}|}$ that satisfy
\begin{align}
\label{eq:polyhedral_TINA_1}
\sum_{s = 1}^{ l_{k} }
d_{k}^{[s]} & \leq \alpha_{kk}^{[l_{k}]}, \
\forall (l_{k},k) \in \mathcal{K} \\
\label{eq:polyhedral_TINA_2}
\sum_{m =1 }^{|\sigma|} \sum_{s = 1}^{l_{\sigma (m)} } d_{\sigma(m)}^{[s]} & \leq
\sum_{m =1 }^{|\sigma|} \alpha_{\sigma(m) \sigma(m) }^{[l_{\sigma(m) }]} - \alpha_{\sigma(m) \sigma(m-1)}^{[l_{\sigma(m) }]}, 
\  \forall l_{\sigma (m)}  \in \langle  L_{\sigma (m)} \rangle ,
 \sigma \in \Sigma.
\end{align}
We denote the above polyhedral region by $\mathcal{P}$.
\end{definition}
As seen in \cite{Joudeh2019a,Joudeh2019b}, several polyhedral-mc-TINA regions can be defined by taking subsets of users, while forcing the GDoF of remaining  users to zero; as well as considering different decoding orders. 
Nevertheless, for the regimes of interest, we only need the polyhedral-mc-TINA region in Definition \ref{def:polyhedral_region}, as seen further on.
Next, we define the  mc-CTIN and mc-TIN regimes, originally identified  in \cite{Joudeh2019a,Joudeh2019b}.
\begin{definition}
\label{def:CTIN_regime}
\textbf{(mc-CTIN Regime).}
In this regime, we have
\begin{align}
\label{eq:CTIN_cond_1}
\alpha_{ii}^{[l_{i}]}  & \geq  \alpha_{ij}^{[l_{i}]} +  \alpha_{ii}^{[l_{i}']}  - \alpha_{ij}^{[l_{i}']}, \
\forall l_{i}',l_{i} \in \langle L_{i} \rangle, \; l_{i}' < l_{i} \\
\label{eq:CTIN_cond_2}
\alpha_{ii}^{[1]} & \geq  \alpha_{ij}^{[1]} + \alpha_{ki}^{[l_{k}]} -
\alpha_{kj}^{[l_{k}]} \mathbbm{1}\big( k \neq j\big), \  \forall l_{k} \in \langle L_k \rangle
\end{align}
for all cells $i,j,k \in \langle K \rangle$, such that  $i \notin \{j,k\}$.
\end{definition}
As shown in \cite{Joudeh2019a},  $\mathcal{D}_{\mathrm{TINA}}^{\mathrm{mc}}$  is a convex polyhedron without the need for time-sharing in the mc-CTIN,
and it coincides with the polyhedral-mc-TIN region in Definition \ref{def:polyhedral_region}, i.e.
\begin{equation}
\label{eq:GDoF_mc_CTIN_regime}
\mathcal{D}_{\mathrm{TINA}}^{\mathrm{mc}} = \mathcal{P}.
\end{equation}
\begin{definition}
\label{def:TIN_regime}
\textbf{(mc-TIN Regime).}
In this regime, we have
\begin{align}
\nonumber
\alpha_{ii}^{[l_{i}]}  & \geq  \alpha_{ij}^{[l_{i}]} +  \alpha_{ii}^{[l_{i}']} 
\ \text{ or }  
\\
\label{eq:TIN_cond_1}
\alpha_{ii}^{[l_{i}]}   & \geq  2\alpha_{ij}^{[l_{i}]} +  \alpha_{ii}^{[l_{i}']}  - \alpha_{ij}^{[l_{i}']}, \
\forall l_{i}',l_{i} \in \langle L_{i} \rangle, \; l_{i}' < l_{i} \\
\label{eq:TIN_cond_2}
\alpha_{ii}^{[1]} & \geq  \alpha_{ij}^{[1]} + \alpha_{ki}^{[l_{k}]}, \  \forall l_{k} \in \langle L_k \rangle
\end{align}
for all cells $i,j,k \in \langle K \rangle$, such that  $i \notin \{j,k\}$.
\end{definition}
The  mc-TIN regime is included in the the mc-CTIN regime.
Moreover, as shown in \cite{Joudeh2019a,Joudeh2019b}, mc-TIN is GDoF optimal for both the IBC ad IMAC in the mc-TIN regime,
that is 
\begin{equation}
\label{eq:GDoF_mc_TIN_regime}
\mathcal{D}^{\mathrm{IBC}} = \mathcal{D}^{\mathrm{IMAC}} = \mathcal{D}_{\mathrm{TINA}}^{\mathrm{mc}}  = \mathcal{P}.
\end{equation}
Note that the results in \cite{Joudeh2019a,Joudeh2019b} are derived without making explicit CSIT assumptions.
Nevertheless, as finite precision CSIT is sufficient to achieve $\mathcal{D}_{\mathrm{TINA}}^{\mathrm{mc}}$, the mc-TIN optimality result 
remains intact under finite precision CSIT.
\begin{figure}[t]
\centering
\includegraphics[width = 0.7\textwidth]{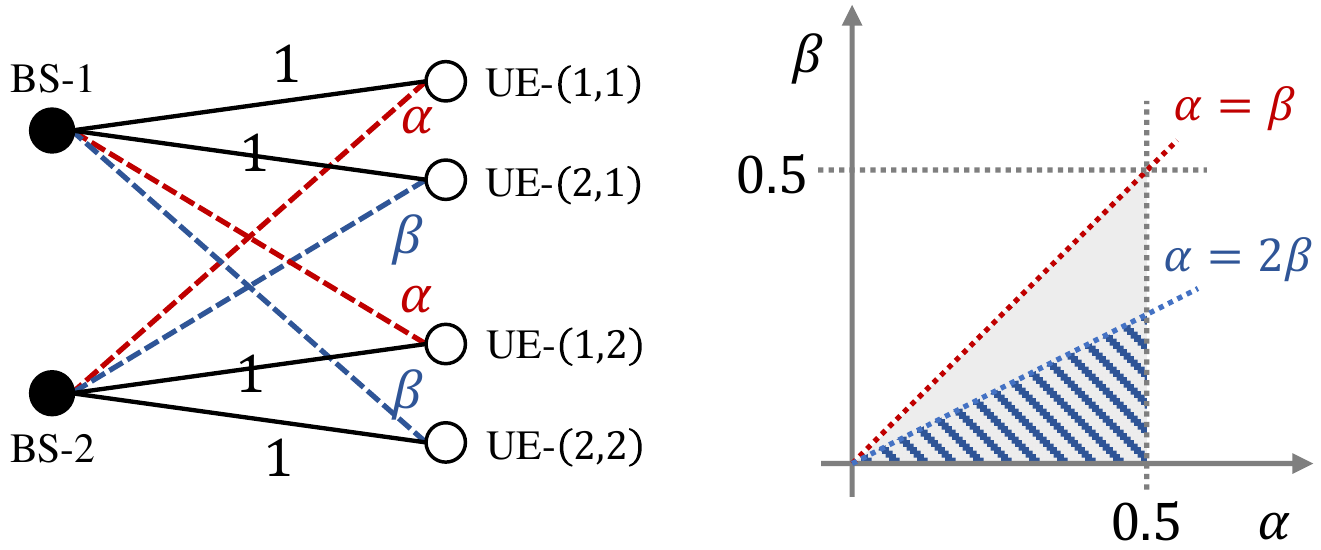}
\caption{Left: symmetric $2$-cell network with $2$ users in each cell ($\alpha \geq \beta$). Right: mc-CTIN regime in grey, and mc-TIN regime in striped blue.}
\label{fig:IBC_IMAC_example}
\end{figure}

The mc-TIN and mc-CTIN regimes are illustrated in Fig. \ref{fig:IBC_IMAC_example} for a simple symmetric network.
Note that in this network, we may assume that $\alpha \geq \beta$ without loss of generality. 
\section{Optimality of mc-TIN in the mc-CTIN Regime} 
\label{sec:main_result}
Here we present the main result of this paper.
\begin{theorem}
\label{theorem:CTIN_optimality}
In the mc-CTIN regime under finite precision CSIT, mc-TIN is 
GDoF optimal for the IBC and IMAC, i.e.
\begin{equation}
\mathcal{D}^{\mathrm{IBC}} = \mathcal{D}^{\mathrm{IMAC}} = 
\mathcal{D}_{\mathrm{TINA}}^{\mathrm{mc}} = \mathcal{P}.
\end{equation}
\end{theorem}
In the mc-CTIN regime, mc-TIN is not necessarily GDoF optimal under perfect CSIT.
This is seen through the example in Fig. \ref{fig:IBC_IMAC_example} for $\alpha = 0.5$ and $\frac{1}{3} < \beta \leq 0.5$, placing this network in the mc-CTIN regime but not in the mc-TIN regime.
In this case, a sum-GDoF of  $\frac{4}{3}$ is achieved through mc-IA \cite{Suh2011,Suh2008},
while the sum-GDoF achieved using mc-TIN is at most $2(1 - \beta) < \frac{4}{3}$.

While mc-IA can surpassed mc-TIN under perfect CSIT, the GDoF regions of the IBC and IMAC, however, remain unknown in general in the mc-CTIN regime.
Theorem \ref{theorem:CTIN_optimality} settles this GDoF region question under finite precision CSIT, and shows that mc-IA has no role to play in this case.
\section{Proof of Theorem \ref{theorem:CTIN_optimality}}
The achievability part of Theorem \ref{theorem:CTIN_optimality}  follows directly from \eqref{eq:GDoF_mc_CTIN_regime}.
To prove the converse part, we wish to show that the polyhedral-mc-TIN region $\mathcal{P}$, described in Definition \ref{def:polyhedral_region}, is a valid outer bound for both the IBC and IMAC in the mc-CTIN regime under finite precision CSIT.

To this end, we work with deterministic approximations of the Gaussian channels.
For the IBC signal model in \eqref{eq:IBC_system model},
the counterpart deterministic model is given by
\begin{equation}
\label{eq:IBC_system model_det}
\bar{Y}_{k}^{[l_{k}]}(t)
 = \sum_{i = 1}^{K} \big\lfloor \bar{P}^{\alpha_{ki}^{[l_{k}]} - \alpha_{\max , i} }  G_{ki}^{[l_{k}]}(t) \bar{X}_{i}(t) 
 \big\rfloor
\end{equation}
where both real and imaginary components of $ \bar{X}_{i}(t) $ are drawn from  
$\big\langle 0 : \lceil \bar{P}^{\alpha_{\max , i}} \rceil  \big\rangle$,
and $ \alpha_{\max , i} \triangleq  \max_{(l_{j} , j)  \in \mathcal{K} } \alpha_{ji}^{[l_{j}]}$.
In the regime of interest, we have $\alpha_{\max , i} = \alpha_{ii}^{[L_{i}]}$.
As shown in \cite{Davoodi2016}, the GDoF of the original channel in \eqref{eq:IBC_system model} is bounded above by the GDoF of the deterministic channel in \eqref{eq:IBC_system model_det}.  

Similarly, the deterministic model corresponding to the 
IMAC  signal model in \eqref{eq:IMAC_system model} is given by
\begin{equation}
\label{eq:IMAC_system model_det}
\bar{Y}_{i}(t)  = \sum_{k = 1}^{K} \sum_{l_{k} = 1}^{L_{k}} \big\lfloor  \bar{P}^{\alpha_{ki}^{[l_{k}]} - \alpha_{k, \max}^{[l_{k}]}  }G_{ki}^{[l_{k}]} \bar{X}_{k}^{[l_{k}]}(t)  \big\rfloor  
\end{equation}
where real and imaginary components of $ \bar{X}_{k}^{[l_{k}]}(t) $ are drawn from  
$\big\langle 0 : \lceil  \bar{P}^{\alpha_{k, \max}^{[l_{k}]} } \rceil  \big\rangle$,
and $ \alpha_{k, \max}^{[l_{k}]}  \triangleq  \max_{ i  \in \langle K \rangle}
\alpha_{k i}^{[l_{k}]}  = \alpha_{k k}^{[l_{k}]} $.
Next, we recall a key lemma from  \cite{Chan2019} (see also  \cite{Davoodi2017a}).
\begin{lemma} \textbf{(Lemma 2.1 \cite{Chan2019})} 
\label{lemma:AI_diff_enropies}
Within the context of the above deterministic model, consider the output signals given by
\begin{align}
\bar{Y}_{k}(t) & =  \sum_{i = 1}^{K} \big\lfloor \bar{P}^{\lambda_{i} - \alpha_{\max,i} } G_{ki}(t) \bar{X}_{i}(t)  \big\rfloor  \\
\bar{Y}_{j}(t) & =  \sum_{i = 1}^{K} \big\lfloor \bar{P}^{\nu_{i} - \alpha_{\max,i} } G_{ji}(t) \bar{X}_{i}(t)  \big\rfloor
\end{align}
where $\lambda_{i},\nu_{i} \in [0, \alpha_{\max,i}]$, for all $i \in \langle K \rangle$, are the corresponding channel strength parameters.
Let $\mathcal{G} \triangleq \{ G_{ki}(t), G_{ji}(t)   :   t \in \langle T \rangle , i \in \langle K \rangle    \}$ be the set of channel coefficients, known to  transmitters up to finite precision. We have 
\begin{equation}
\label{eq:entropy_diff_lemma}
H\big( \bar{\mathbf{Y}}_{k} | \mathcal{G}, U \big) - 
H\big( \bar{\mathbf{Y}}_{j} | \mathcal{G}, U \big) \leq  
\max_{i \in \langle K \rangle} 
(\lambda_{i} - \nu_{i})^{+} T \log(P) + T o(\log(P))
\end{equation}
where $U$ is an auxiliary random variable, independent of $\mathcal{G}$.
\end{lemma}
The above lemma bounds the maximum difference of entropies (in the GDoF sense) between the two received signals 
$ \bar{\mathbf{Y}}_{k} $ and $ \bar{\mathbf{Y}}_{j}$, that can be created by any set of feasible input signals 
$\bar{\mathbf{X}}_{1}, \ldots, \bar{\mathbf{X}}_{K}$,
which are independent of the exact realizations of channel coefficients 
$\mathcal{G}$.
The bound in \eqref{eq:entropy_diff_lemma} tells us that a maximal difference of entropies, in the GDoF sense, is created through $\bar{\mathbf{X}}_{i}$, where $i$ is the index yielding a maximum difference in strengths $(\lambda_{i} - \nu_{i})^{+}$.
For more insights related to this lemma, readers are referred to \cite{Chan2019}  and  \cite{Davoodi2017a}.

For brevity, we adopt the compact notation of \cite{Chan2019} to represent differences of entropies of the type  in  Lemma \ref{lemma:AI_diff_enropies}.
Using this compact notion,  inequality \eqref{eq:entropy_diff_lemma} is expressed as
\begin{equation}
\label{eq:entropy_diff_lemma_brief}
\Hbb \big( [\lambda_{1} , \ldots, \lambda_{K}] \big)  -  \Hbb \big( [\nu_{1} , \ldots, \nu_{K}]\big)  \leq   \max_{i \in \langle K \rangle} 
(\lambda_{i} - \nu_{i})^{+} T \log(P)
\end{equation}
where the $o(\log(P))$ term is  dropped as it is inconsequential for GDoF purposes (we ignore such terms henceforth).
The expression in \eqref{eq:entropy_diff_lemma_brief} succinctly captures the essential parts in \eqref{eq:entropy_diff_lemma}, 
i.e. channel strength levels, and will be employed further on.
We are now fully equipped to derive converse bounds for the IBC and IMAC under finite precision CSIT.
\subsection{Converse for the IBC}
The single-cell bounds in \eqref{eq:polyhedral_TINA_1} are trivially obtained from the capacity region of the degraded Gaussian BC, see, e.g., \cite{Cover2012}.
Therefore, we focus on the multi-cell bounds in  \eqref{eq:polyhedral_TINA_2}.
In what follows, $\mathcal{G}$ denotes the set of all channel coefficients.

Consider the cycle $\sigma \in \Sigma$ of length $|\sigma | = M$.
Moreover, consider $(l _{\sigma(1)} , \ldots, l_{\sigma(M)} ) \in \langle L _{\sigma(1)} \rangle \times  \ldots \times \langle L_{\sigma(M)} \rangle$, which determines the number and identifies of participating users from each participating cell.
Let us first consider a single participating cell $i \in \{ \sigma \}$ and its $l_{i}$ participating users, and focus on the corresponding  sum-rate 
$\sum_{s_{i} = 1}^{l_{i}  } \  R_{i}^{[s_{i}]}$.

From Fano's inequality, we obtain
\begin{align}
\nonumber
 T &  \sum_{s = 1}^{l_{i}  } \  R_{i}^{[s]} \leq \
 \sum_{s = 1}^{l_{i}}    I \Big(  W_{i}^{[s]} ;  \bar{\mathbf{Y}}_{i}^{[s]}  | \mathcal{G},  W_{i}^{[1:s - 1]}   \Big) \\
 \nonumber
 = &  \sum_{s = 1}^{l_{i}}   H  \Big(   \bar{\mathbf{Y}}_{i}^{[s]}  |   \mathcal{G},W_{i}^{[1:s - 1]}  \Big)  - 
 H \Big(  \bar{\mathbf{Y}}_{i}^{[s]}  |  \mathcal{G}, W_{i}^{[1:s ]}  \Big)  \\
\label{eq:converse_IBC_single_cell_1}
  = &   \sum_{s = 2}^{l_{i}}   \Big[ H  \Big(   \bar{\mathbf{Y}}_{i}^{[s ]}  |  \mathcal{G}, W_{i}^{[1:s - 1]}  \Big)  - 
 H \Big(  \bar{\mathbf{Y}}_{i}^{[s - 1]}  |  \mathcal{G}, W_{i}^{[1:s - 1 ]}  \Big)   \Big]
+ H  \Big(   \bar{\mathbf{Y}}_{i}^{[1]} |  \mathcal{G}  \Big)  - H \Big(  \bar{\mathbf{Y}}_{i}^{[l_{i}]}  |  \mathcal{G}, W_{i}^{[1:l_{i}]}  \Big)  \\
\label{eq:converse_IBC_single_cell_2}
  = &   \sum_{s = 2}^{l_{i}}   \!   \Big[  \Hbb \Big( \big[ \alpha^{[s ]}_{i \sigma(1)} , \ldots, \alpha^{[s ]}_{i \sigma(M)}  \big] \Big)  \! -   \! 
  \Hbb \Big(  \big[ \alpha^{[s -1 ]}_{i \sigma(1)} , \ldots, \alpha^{[s  -1 ]}_{i \sigma(M)}  \big]  \Big)   \Big]
  + H  \Big(   \bar{\mathbf{Y}}_{i}^{[1]} |  \mathcal{G}  \Big)  - H \Big(  \bar{\mathbf{Y}}_{i}^{[l_{i}]}  |  \mathcal{G}, W_{i}^{[1:l_{i}]}  \Big)
\end{align}
where in going from \eqref{eq:converse_IBC_single_cell_1} to \eqref{eq:converse_IBC_single_cell_2}, we have invoked the brief notation in \eqref{eq:entropy_diff_lemma_brief}.
Now let us focus on bounding the sum of differences of entropies in the brief notation in \eqref{eq:converse_IBC_single_cell_2}.
Using the result in Lemma \ref{lemma:AI_diff_enropies}, we obtain 
\begin{align}
\nonumber
& \sum_{s = 2}^{l_{i}}   \!    \Hbb \Big( \big[ \alpha^{[s ]}_{i \sigma(1)} , \ldots, \alpha^{[s  ]}_{i \sigma(M)}  \big] \Big)  \! -   \! 
  \Hbb \Big(  \big[ \alpha^{[s -1 ]}_{i \sigma(1)} , \ldots, \alpha^{[s -1 ]}_{i \sigma(M)}  \big]  \Big)  \\
\label{eq:converse_IBC_single_cell_3}
&\leq \sum_{s = 2}^{l_{i}}  \max_{j \in \{ \sigma \} } \left(  \alpha^{[s ]}_{ij}  - \alpha^{[s  -1 ]}_{ij}   \right)^{+}  T \log (P) \\ 
\label{eq:converse_IBC_single_cell_4}
& \leq  \sum_{s = 2}^{l_{i}}  \left(    \alpha^{[s ]}_{ii}  - \alpha^{[s -1 ]}_{ii} \right) T \log (P)  \\
\label{eq:converse_IBC_single_cell_5}
& =  \left(    \alpha^{[l_{i} ]}_{ii}  - \alpha^{[1]}_{ii} \right) T \log (P).
\end{align}
Going from \eqref{eq:converse_IBC_single_cell_3} to \eqref{eq:converse_IBC_single_cell_4} follows from the mc-CTIN condition in \eqref{eq:CTIN_cond_1}, which implies
\begin{equation}
\nonumber
\alpha^{[s ]}_{ii} - \alpha^{[s  ]}_{ij}  \geq  \alpha^{[s - 1 ]}_{ii} - \alpha^{[s -1  ]}_{ij}   \implies 
\alpha^{[s ]}_{ii} - \alpha^{[s - 1 ]}_{ii}   \geq   \alpha^{[s  ]}_{ij}  - \alpha^{[s -1  ]}_{ij}, \ \forall s  \in \langle 2 : l_{i} \rangle, 
\end{equation}
as well as the order $\alpha_{ii}^{[s]} \geq \alpha_{ii}^{[s-1]}$ in \eqref{eq:strength_order}, which allows us to drop the $(\cdot)^{+}$.
By combining  \eqref{eq:converse_IBC_single_cell_2} and \eqref{eq:converse_IBC_single_cell_5}, we obtain
\begin{equation}
\label{eq:converse_IBC_single_cell_final}
 T  \sum_{s = 1}^{l_{i}  } \  R_{i}^{[s ]} \leq  \left(    \alpha^{[l_{i} ]}_{ii}  - \alpha^{[1]}_{ii} \right) T \log (P) 
+ H  \Big(   \bar{\mathbf{Y}}_{i}^{[1]} |  \mathcal{G}  \Big)  - H \Big(  \bar{\mathbf{Y}}_{i}^{[l_{i}]}  |  \mathcal{G}, \bar{\mathbf{X}}_{i}  \Big)
\end{equation}
which holds for any single cell $i \in \{ \sigma \}$.
Next, we combine the bounds obtained from \eqref{eq:converse_IBC_single_cell_final}, for all $i \in \{ \sigma \}$,
to obtain multi-cell sum-rate bound for participating users as 
\begin{align}
\nonumber
& T \sum_{m = 1}^{M}  \sum_{s = 1}^{l_{\sigma(m)}  }  R_{\sigma (m)}^{[s]} \leq 
\sum_{m = 1}^{M}   \!  \!  \left(    \alpha^{[l_{\sigma(m)} ]}_{\sigma(m)\sigma(m)}  - 
\alpha^{[1]}_{\sigma(m)\sigma(m)} \right) T \log (P) \\
\label{eq:converse_IBC_multi_cell_1}
& + \sum_{m = 1}^{M} H  \Big(   \bar{\mathbf{Y}}_{\sigma(m)}^{[1]} |  \mathcal{G}  \Big)  - H \Big(  \bar{\mathbf{Y}}_{\sigma(m)}^{[l_{\sigma(m)}]}  |  \mathcal{G}, \bar{\mathbf{X}}_{\sigma(m)}  \Big).
\end{align}
Focusing on the differences of entropies in  \eqref{eq:converse_IBC_multi_cell_1},
we obtain
\begin{align}
\nonumber
 \sum_{m = 1}^{M} & H  \Big(   \bar{\mathbf{Y}}_{\sigma(m)}^{[1]} |  \mathcal{G}  \Big)  - H \Big(  \bar{\mathbf{Y}}_{\sigma(m+1)}^{[l_{\sigma(m+1)}]}  |  \mathcal{G}, \bar{\mathbf{X}}_{\sigma(m+1)}  \Big) \\
\label{eq:converse_IBC_multi_cell_2}
= & \!  \!  \sum_{m = 1}^{M}  \! 
\Hbb \!  \Big( \!  \big[  \alpha^{[1]}_{\sigma(m)\sigma(1)} , \ldots, 
\alpha^{[1]}_{\sigma(m)\sigma(m+1)}  ,\ldots, 
\alpha^{[1 ]}_{\sigma(m) \sigma(M)}  \big] \!  \Big)   
\! -  \! \Hbb \Big( \big[ \alpha^{[l_{\sigma(m+1)}]}_{\sigma(m+1)\sigma(1)} , \ldots,  0 ,\ldots, 
\alpha^{[l_{\sigma(m+1)}]}_{\sigma(m+1) \sigma(M)}  \big] \Big)  \\
\label{eq:converse_IBC_multi_cell_pre3}
\leq  &  \ T \log (P)  \sum_{m = 1}^{M} \max \bigg(  \alpha^{[1]}_{\sigma(m)\sigma(m+1)} ,  
\max_{j \in \langle M \rangle, j \neq m+1} \left(  \alpha^{[1]}_{\sigma(m)\sigma(j)} 
- \alpha^{[l_{\sigma(m+1)}]}_{\sigma(m+1)\sigma(j)}   \right)^{+}  \bigg) \\
\label{eq:converse_IBC_multi_cell_3}
 \leq  & \  T \log (P)  \sum_{m = 1}^{M}   \left(   \alpha^{[1]}_{\sigma(m)\sigma(m)}   - \alpha^{[l_{\sigma(m+1)}]}_{\sigma(m+1)\sigma(m)}  \right).
\end{align}
In each of the negative entropy terms on the left-hand-side of \eqref{eq:converse_IBC_multi_cell_2},
the contribution of the input signal $\bar{\mathbf{X}}_{\sigma(m+1)} $ can be removed from the output signal 
$\bar{\mathbf{Y}}_{\sigma(m+1)}^{[l_{\sigma(m+1)}]} $, after which the conditioning on $\bar{\mathbf{X}}_{\sigma(m+1)} $  can be dropped
since different input signals are independent.
This leads to the right-hand-side of \eqref{eq:converse_IBC_multi_cell_2}, 
where in each negative entropy term,  the signal power level corresponding to $\bar{\mathbf{X}}_{\sigma(m+1)} $  is zeroed. 
On the other hand, \eqref{eq:converse_IBC_multi_cell_pre3} is obtained from a direct application of Lemma \ref{lemma:AI_diff_enropies}.

The bound in  \eqref{eq:converse_IBC_multi_cell_3} holds due to the
 mc-CTIN conditions.
In particular, \eqref{eq:CTIN_cond_2} implies that for $j \neq m+1$, we have
\begin{align}
\nonumber
\alpha^{[1]}_{\sigma(m)\sigma(m)}   - \alpha^{[l_{\sigma(m+1)}]}_{\sigma(m+1)\sigma(m)}  &  \geq  \alpha^{[1]}_{\sigma(m)\sigma(j)} 
- \alpha^{[l_{\sigma(m+1)}]}_{\sigma(m+1)\sigma(j)}  \\
\nonumber
\alpha^{[1]}_{\sigma(m)\sigma(m)}   - \alpha^{[l_{\sigma(m+1)}]}_{\sigma(m+1)\sigma(m)}  & \geq \alpha^{[1]}_{\sigma(m)\sigma(m+1)}  
\end{align}
from which \eqref{eq:converse_IBC_multi_cell_3} directly follows.
By combining  the bounds in  \eqref{eq:converse_IBC_multi_cell_3}  and \eqref{eq:converse_IBC_multi_cell_1},
we obtain the desired cycle bound as
\begin{align}
\nonumber
T \sum_{m = 1}^{M}  \sum_{s = 1}^{l_{\sigma(m)}  }  R_{\sigma (m)}^{[s]}  
& \leq  \sum_{m = 1}^{M}   \left(
    \alpha^{[l_{\sigma(m)}]}_{\sigma(m)\sigma(m)}   - 
 \alpha^{[l_{\sigma(m+1)}]}_{\sigma(m+1)\sigma(m)}  \right) T \log (P) \\ 
&  =  \sum_{m = 1}^{M}   \left(
    \alpha^{[l_{\sigma(m)}]}_{\sigma(m)\sigma(m)}   - 
 \alpha^{[l_{\sigma(m)}]}_{\sigma(m)\sigma(m-1)}  \right) T \log (P).
\end{align}
This concludes the converse proof for the IBC.
\subsection{Converse for the IMAC}
Considering the the same  $\sigma$ and $(l _{\sigma(1)} , \ldots, l_{\sigma(M)} )$ for the IMAC,
and starting from Fano's inequality, we have the following sequence of bounds on the cycle sum-rate
\begin{align}
\nonumber
T & \sum_{m = 1}^{M}  \sum_{s = 1}^{l_{\sigma(m)}  }  R_{\sigma (m)}^{[s]} \leq  \sum_{m = 1}^{M}  
I \Big(  \bar{\mathbf{X}}_{\sigma(m)}^{[1,\ldots, l_{\sigma(m)} ]}  ;  \bar{\mathbf{Y}}_{\sigma(m)}  |  \mathcal{G}  \Big)  \\
\nonumber 
= & \sum_{m = 1}^{M}   H \Big( \bar{\mathbf{Y}}_{\sigma(m)}   |  \mathcal{G}  \Big)  
- H \Big(   \bar{\mathbf{Y}}_{\sigma(m)}  |  \mathcal{G} , \bar{\mathbf{X}}_{\sigma(m)}^{[1,\ldots, l_{\sigma(m)} ]}     \Big)  \\
\nonumber
 = & \sum_{m = 1}^{M}   H \Big( \bar{\mathbf{Y}}_{\sigma(m)} |  \mathcal{G}   \Big)  
- H \Big(   \bar{\mathbf{Y}}_{\sigma(m-1)}  |  \mathcal{G} , \bar{\mathbf{X}}_{\sigma(m-1)}^{[1,\ldots, l_{\sigma(m-1)} ]}     \Big)  \\
\label{eq:converse_IMAC_1}
= & \sum_{m = 1}^{M} \!    \Hbb \Big(  \!  \big[ 
\alpha^{[1 : l_{\sigma(1)}]}_{\sigma(1)\sigma(m)} , \ldots,  \alpha^{[1 : l_{\sigma(m-1)}]}_{\sigma(m-1)\sigma(m)}    ,\ldots, 
\alpha^{[1 : l_{\sigma(M)}]}_{\sigma(M)\sigma(m)}  \big]   \!  \Big) 
 - \Hbb \Big(  \!  \big[ 
\alpha^{[1 : l_{\sigma(1)}]}_{\sigma(1)\sigma(m-1)} , \ldots,  \mathbf{0}   ,\ldots, 
\alpha^{[1 : l_{\sigma(M)}]}_{\sigma(M)\sigma(m-1)}  \big]   \!  \Big)  \\
\nonumber
 \leq  & \ T \log (P) \sum_{m = 1}^{M} \max \bigg( \max_{s \in \langle l_{\sigma(m-1)} \rangle }  \left(   
\alpha^{[s]}_{\sigma(m-1)\sigma(m)}   \right) ,  
\max_{j \in \langle M \rangle, j \neq m-1 }  \max_{s \in \langle l_{\sigma(j)} \rangle } 
\left( \alpha^{[s]}_{\sigma(j)\sigma(m)}  - \alpha^{[s]}_{\sigma(j)\sigma(m-1)}  \right)^{+} \bigg) \\
\label{eq:converse_IMAC_2}
\leq  & \ T \log (P) \sum_{m = 1}^{M} \left( \alpha^{[l_{\sigma(m)}]}_{\sigma(m)\sigma(m)}   - \alpha^{[l_{\sigma(m)}]}_{\sigma(m)\sigma(m-1)}  \right).
\end{align}
In the above, \eqref{eq:converse_IMAC_1} is obtained by employing the same argument used for \eqref{eq:converse_IBC_multi_cell_2} in the previous part,
where $\mathbf{0}$ is used to denote a sequence of zeros of  length $l_{\sigma(m-1)}$.
The following inequality is obtained by applying Lemma \ref{lemma:AI_diff_enropies} to \eqref{eq:converse_IMAC_1}.
The inequality in \eqref{eq:converse_IMAC_2} holds due to the mc-CTIN conditions.
In particular,  
for all $j \neq m- 1$ and $s \in  \langle l_{\sigma(j)} \rangle$, we have 
\begin{align}
\label{eq:converse_IMAC_CTIN_cond_1}
\alpha^{[l_{\sigma(m)}]}_{\sigma(m)\sigma(m)}    \!  - \alpha^{[l_{\sigma(m)}]}_{\sigma(m)\sigma(m-1)}  
& \geq  \alpha^{[1]}_{\sigma(m)\sigma(m)}  \!  \! -  \! \alpha^{[1]}_{\sigma(m)\sigma(m-1)}   \\
\label{eq:converse_IMAC_CTIN_cond_2}
& \geq \alpha^{[s]}_{\sigma(j)\sigma(m)} 
 \!  \! - \alpha^{[s]}_{\sigma(j)\sigma(m-1)}    
\end{align}
where \eqref{eq:converse_IMAC_CTIN_cond_1} holds due to \eqref{eq:CTIN_cond_1}, while 
\eqref{eq:converse_IMAC_CTIN_cond_2} holds due to \eqref{eq:CTIN_cond_2}.
Moreover, the mc-CTIN condition in \eqref{eq:CTIN_cond_2}  further implies 
\begin{align}
\label{eq:converse_IMAC_CTIN_cond_3}
\alpha^{[l_{\sigma(m)}]}_{\sigma(m)\sigma(m)}   - \alpha^{[l_{\sigma(m)}]}_{\sigma(m)\sigma(m-1)}  & \geq \alpha^{[s]}_{\sigma(m-1)\sigma(m)}
\end{align}
for all $s \in  \langle l_{\sigma(m-1)} \rangle$.
It can be verified that \eqref{eq:converse_IMAC_CTIN_cond_2} and \eqref{eq:converse_IMAC_CTIN_cond_3} suffice to obtain 
\eqref{eq:converse_IMAC_2} from its preceding inequality. 
This concludes the converse proof for the IMAC, and the proof of Theorem \ref{theorem:CTIN_optimality}.
\section{Conclusion}
In this paper, we  characterized  the GDoF regions of the IBC and IMAC in the mc-CTIN regime under the assumption of finite precisions CSIT; and showed that in this case, mc-TIN is GDoF optimal.
The proof of our results is based on a new application of robust outer bounds based on the aligned images approach.
By following in the footsteps of Chan et al. \cite{Chan2019}, the mc-TIN optimality result under finite precision CSIT here  can serve as a first step towards a multi-cell extremal network study, where robust gains of multi-cell cooperation over multi-cell TIN are investigated in various  regimes of interest.

\bibliographystyle{IEEEtran}
\bibliography{References}
\end{document}